\documentclass[letterpaper,final,notitlepage]{iopart}
\usepackage{iopams} 
\usepackage{graphicx}
\usepackage{color}
\usepackage{epsfig}
\usepackage{times}
\usepackage{cite}

\renewcommand{\vec}{\mathbf}
\newcommand{\leftexp}[2]{{\vphantom{#2}}^{#1}{#2}}

\begin{document}

\title{Quantum decoherence in the rotation of small molecules}

\author{A.~Adelsw\"ard and S.~Wallentowitz\footnote[7]{email:
    sascha.wallentowitz@physik.uni-rostock.de}} 

\address{Emmy--Noether Nachwuchsgruppe ``Kollektive Quantenmessung
  und R\"uckkopplung an Atomen und Molek\"ulen'', Fachbereich Physik,
  Universit\"at Rostock, Universit\"atsplatz 3, D-18051 Rostock,
  Germany}

%\date{\today}

\begin{abstract}
  The dynamics of non-polar diatomic molecules interacting with a far-detuned
  narrow-band laser field, that only may drive rotational transitions, is
  studied. The rotation of the molecule is considered both classically and
  quantum mechanically, providing links to features known from the heavy
  symmetric top. In particular, quantum decoherence in the molecular rotation,
  being induced by spontaneous Raman processes, is addressed. It is shown how
  this decoherence modifies the rotational dynamics in phase space.
\end{abstract}


% 03.65.Yz Decoherence; open systems; quantum statistical methods
% 33.80.-b Photon interactions with molecules (see also 42.50 Quantum
% optics)
% 33.80.Ps Optical cooling of molecules; trapping
% 42.50.Ct Quantum description of interaction of light and matter;
% related experiments
% 42.50.Lc Quantum fluctuations, quantum noise, and quantum jumps

\section{Introduction} \label{sec:1}
Molecules interacting with electromagnetic fields foster a variety of
dynamical phenomena. Specific wave packets of the internuclear vibration have
been excited by shaped fs laser pulses~\cite{fs-pulse}. Such vibrational
quantum states have then been reconstructed by molecular emission
tomography~\cite{met1,met2}. Furthermore, control of the internal molecular
quantum state could be achieved~\cite{ctrl1,ctrl2,ctrl3,ctrl4,ctrl5}, for
eventually enhancing chemical reactions.

Besides these applications of pulsed fields, also interactions with cw laser
fields or static fields reveal interesting effects. For instance, the axis of
paramagnetic molecules can be aligned in pendular states by applying magnetic
fields~\cite{herschbach-B1,herschbach-B2}.  Furthermore, it has been shown how
electric fields align polar molecules~\cite{herschbach-E}.  Non-polar
molecules, on the other hand, seem to be rather isolated from the influence of
electric fields. They are infrared inactive and thus their rotation is
essentially undamped and isolated from the electromagnetic vacuum. Clearly
this does not hold for the internuclear vibration and, in addition, the
rotational statistics determines via ro-vibrational coupling its relaxation
and decoherence properties~\cite{wal-mol1,wal-mol2}.

Nevertheless, a permanent dipole moment being absent, non-polar molecules
still can be polarised by electrical fields and thus feel an effective
alignment force when an anisotropic polarisability prevails. The
quantum-mechanical eigenstates and energies have been discussed in
Ref.~\cite{herschbach-pol}, where it has been shown that the dependence of the
eigenenergies on the field intensity can be used for providing a dipole
trapping force. Recent experiments have implemented such a dipole trap for
molecules~\cite{PA-trap-dipole}.

In general the topic of
producing~\cite{mol-PA1,mol-PA2,mol-PA3,mol-PA-BEC1,mol-PA-BEC2,mol-PA-BEC3},
cooling~\cite{buffergas,e-field}, and trapping of ultra-cold
molecules~\cite{PA-trap-dipole,mol-trap-magn} became a major focus of research
in recent years.  Given the ultra-low temperatures of a condensed atomic or
possibly molecular gas, the molecular rotation, representing the lowest energy
scale apart from centre-of-mass motion, then should play an important role.
For sufficiently cold molecules one might expect decoherence first to appear
on the rotational energy scale before affecting the energetically higher
degrees of freedom, such as the internuclear vibration. Moreover, the
molecular rotation provides a toolbox for implementing various textbook
examples of mechanics, such as spherical pendula and the motion of rigid
bodies.

In particular, we will focus here on homonuclear, diatomic molecules in a
far-detuned linear-polarised laser field.  This is a setup relevant also for
trapping of photo-associated ultra-cold molecules in a dipole
trap~\cite{PA-trap-dipole}. In Sec.~\ref{sec:2} a brief review is given on the
classical description of the motion of the molecular axis. A link to a quantum
description in terms of stimulated two-photon Raman processes is then
presented in Sec.~\ref{sec:3}. Moreover, the theory is extended there by
including spontaneous Raman processes in form of a quantum master equation.
The effects due to these spontaneous processes on the molecular rotation are
evaluated in Sec.~\ref{sec:4} where we provide a phase-space picture by using
a Wigner function for the molecular rotation.

\section{Classical mechanical analogue}  
\label{sec:2}

\subsection{Classical dynamics} \label{sec:2.1}
\begin{figure}
  \centering
  \epsfig{file=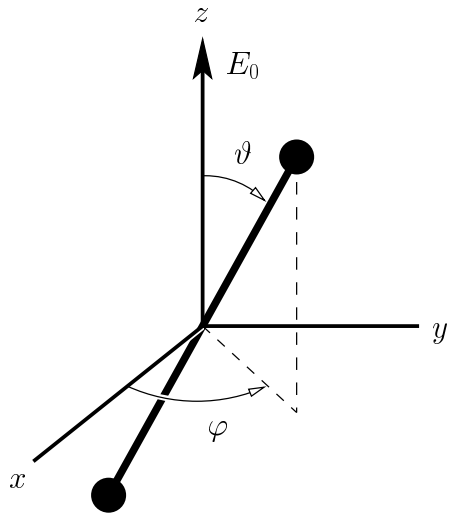,scale=0.5}
  \caption{Diatomic molecule in an electric field pointing in $z$
  direction. The angles $\vartheta$ and $\varphi$ specify the orientation of
  the molecular axis with respect to the laboratory frame.}
  \label{fig:geometry}
\end{figure}

Consider a diatomic molecule in an electric field of amplitude $E_0$ pointing
towards the $z$ direction. The potential energy in such a field is given by
\begin{equation}
  \label{eq:pot-erg}
  V(\vartheta) = -d E_0 \cos\vartheta - {\textstyle\frac{1}{2}}
  E_0^2 \Delta \alpha
  \cos^2\!\vartheta  , 
\end{equation}
where $d$ is the permanent dipole moment of the molecule and $\Delta\alpha
\!=\!  \alpha_\parallel \!-\! \alpha_\perp$ denotes the anisotropy of the
polarisability with respect to fields parallel and perpendicular to the
molecular axis. The direction of this axis with respect to the electric field
is specified by the enclosing angle $\vartheta$, cf.~Fig.~\ref{fig:geometry}.

The molecular axis specifies the $z'$-axis of a body-fixed coordinate system,
whose orientation with respect to the laboratory frame is specified by Euler
angles.  From the Lagrange equations one obtains the equations of motion for
these as
\begin{eqnarray}
  \label{eq:motion-euler}
  \Theta \big( \ddot{\vartheta} - \dot{\varphi}^2 \sin\vartheta \cos\vartheta
  \big) + \frac{\partial V(\vartheta)}{\partial \vartheta} = 0 , \\
  \label{eq:Lz-const}
  \dot{\varphi} \, \sin^2\!\vartheta = {\rm const} , 
\end{eqnarray}
where $\Theta$ is the moment of inertia perpendicular to the molecular axis.
Moreover, the conservation of energy $E$ reads
\begin{equation}
  \label{eq:E-const}
   \Theta \big( \dot{\vartheta}^2 + \dot{\varphi}^2 \sin^2\!\vartheta \big) 
   = 2 \left[ E \!-\! V(\vartheta) \right]  .
\end{equation}
Note that for our case the relevant Euler angles $\vartheta$ and $\varphi$
coincide with the spherical angles as depicted in Fig.~\ref{fig:geometry}.

The angular-momentum $\vec{L}$ can be expressed in the laboratory frame by the
angles and their time derivatives as
\begin{equation}
  \label{eq:L-vec}
  \vec{L} = \Theta \left( \begin{array}{c}
      \dot{\vartheta} \, \cos\varphi - \dot{\varphi} \, 
      \sin\vartheta \cos\vartheta \sin\varphi \\ 
      \dot{\vartheta} \, \sin\varphi + \dot{\varphi} \, 
      \sin\vartheta \cos\vartheta \cos\varphi \\ 
      \dot{\varphi} \, \sin^2\!\vartheta
    \end{array} \right) .
\end{equation}
From Eqs~(\ref{eq:Lz-const}) and (\ref{eq:L-vec}) it is seen that the
component of the angular momentum in direction of the applied field is a
constant of motion: $L_z \!=\! {\rm const}$. Thus the tip of the
angular-momentum moves in a plane normal to the electric-field direction.
Using the constant of motion $L_z$, Eq.~(\ref{eq:E-const}) can be rewritten as
\begin{equation}
  \frac{\Theta \dot{\vartheta}^2}{2} = E - U(\vartheta) ,
\end{equation}
where the new potential reveals a centrifugal-type barrier
\begin{equation}
  U(\vartheta) = \frac{T_z}{
  \sin^2\!\vartheta} - U_d \cos\vartheta - U_\alpha \cos^2\!\vartheta .
\end{equation}
For later purposes we have introduced here the potential depths $U_d \!=\!
dE_0$, $U_\alpha \!=\!  \Delta \alpha E_0^2/ 2$ and the $z$-part of
the kinetic energy $T_z \!=\! L_z^2 / (2\Theta)$.

\begin{figure}
  \begin{center}
    \epsfig{file=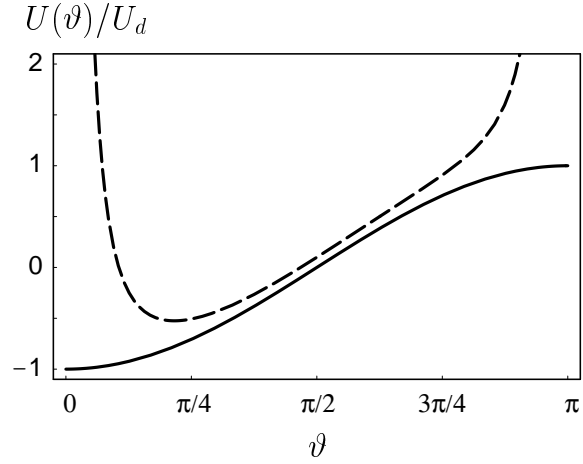,scale=0.75}
  \end{center}
  \caption{Nutation potential $U(\vartheta)$ for a polar molecule scaled by
    the potential depth $U_d$. Solid (dashed) curve corresponds to $T_z / U_d
    \!=\! 0$ $(0.1)$.}
  \label{fig:d-potential}
\end{figure}

A molecule with permanent dipole moment, $d \!\neq\! 0$, omitting its
anisotropic polarisability, $\alpha_\parallel \!\approx\!  \alpha_\perp$,
reveals a potential of a spherical pendulum, as shown in
Fig.~\ref{fig:d-potential}.
The motion in $\varphi$ with constant angular momentum $L_z$ is superimposed
by a nutation in $\vartheta$, which is equivalently described by a heavy
symmetrical top with vanishing moment of inertia along the (molecular) body
axis. The harmonically approximated nutation frequency is in this case $\omega
\!=\!  \sqrt{U_d/ \Theta}$.

\begin{figure}
  \begin{center}
    \epsfig{file=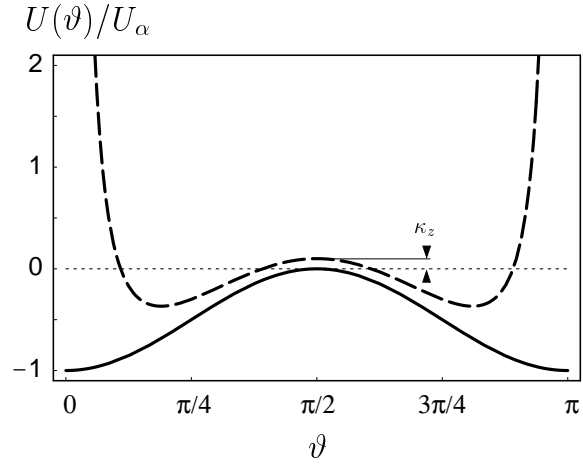,scale=0.75}
  \end{center}
  \caption{Nutation potential $U(\vartheta)$ for a nonpolar, polarisable
    molecule scaled by $U_\alpha$. Solid (dashed) curve corresponds to
    $\kappa_z \!=\! T_z / U_\alpha \!=\! 0$ $(0.1)$.}
  \label{fig:alpha-potential}
\end{figure}
However, for the case of non-polar ($d \!=\! 0$) but polarisable molecules
($\alpha_\parallel \!>\! \alpha_\perp$), which is the focus of this paper, the
anisotropic polarisability produces a $\cos^2\!\vartheta$ potential,
cf.~Fig.~\ref{fig:alpha-potential}, which for $L_z \!\neq\! 0$ cannot be cast
into the form of a spherical pendulum.  Nevertheless, a
spherical-pendulum-type motion can be expected, since also for this case the
solution $\vartheta(t)$ can be obtained in terms of elliptic integrals.
Expressed in terms of Jacobian elliptic functions~\cite{as} it reads,
\begin{equation}
  \label{eq:theta-solution}
  \fl \cos\vartheta(t)  = \frac{ \cos \vartheta_0 \, {\rm cn}(
  \omega t, m ) - \sqrt{(\cos^2\!\vartheta_0 \!-\! c_{l}) (c_{u}
  \!-\! \cos^2 \! \vartheta_0) \, m / c_u } 
  \, {\rm sn}( \omega t, m ) \, {\rm dn}(\omega t, m)}{1 - 
  (c_u \!-\! \cos^2 \!
  \vartheta_0)  \, {\rm sn}^2 (\omega t, m) \, m/ c_u} , 
\end{equation}
where $\vartheta_0$ is the initial angle at time $t \!=\! 0$. The constants
appearing in Eq.~(\ref{eq:theta-solution}) are given as
\begin{eqnarray}
  \label{eq:alpha}
   m  & = &  \frac{1}{\epsilon \!+\! 1} , \\
   c_{l,u} & = & \frac{m \!-\! \frac{1}{2}}{m} \mp \sqrt{ \frac{1}{4m^2}
 - \kappa_z} , \\
  \omega & = & \sqrt{\frac{2U_\alpha}{\Theta} \frac{c_u}{m}} ,
\end{eqnarray}
where the scaled energies $\epsilon \!=\! E/U_\alpha$ and $\kappa_z \!=\!
T_z/U_\alpha$ have been used.

\subsection{Classical types of motion} \label{sec:2.2}
In the asymptotic limit of large energies $\epsilon \!\gg\! 1$, ${\rm sn},
{\rm cn} \!\to\! \sin, \cos$ and ${\rm dn} \!\to\! 1$ so that
$\cos\vartheta(t) \!=\! \cos(\vartheta_0 \!+\! \omega t)$ results, which
corresponds to the free rotator case with angular velocity $\omega \!\to\!
\sqrt{2 E / \Theta}$. In this limit all three vector components of the angular
momentum are conserved quantities.

\begin{figure}
  \centering
  \epsfig{file=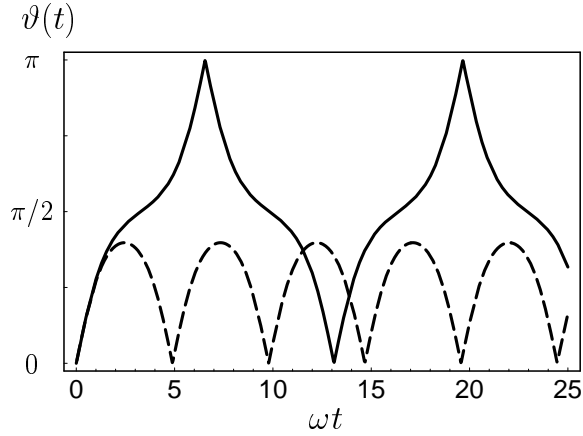, scale=0.75}
  \caption{Nutation angle $\vartheta$ over scaled time $\omega t$ for $\kappa_z
    \!=\! 0$, $\vartheta_0 \!=\! 0$, $L_y(0)\!=\! 0$ and $2 \Theta U_\alpha /
    \hbar^2 \!=\!  0.025$.  Scaled energies and $x$-components of angular
    momentum are: $\epsilon \!=\! 0.024$, $L_x(0) / \hbar \!=\! 0.16$ (solid
    curve); $\epsilon \!=\! -0.1$, $L_x(0) / \hbar \!=\! 0.15$ (dashed
    curve).}
  \label{fig:kappaz=0}
\end{figure}

\begin{figure}
  \centering
  \epsfig{file=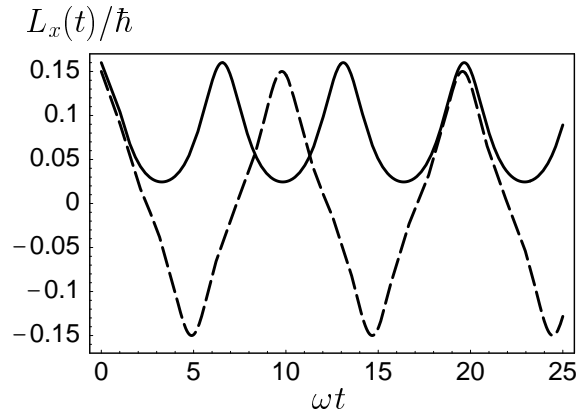, scale=0.75}
  \caption{$L_x / \hbar$ over scaled time $\omega t$ for the same parameters
    as in Fig.~4.}
  \label{fig:Ly-kappaz=0}
\end{figure}

For lower energies the potential barriers lead to modifications, supporting
several types of motion. For $\kappa_z \!=\! 0$ unbound motion occurs for
$\epsilon \!>\! 0$ where the amplitude of the oscillation of $\vartheta$
attains the full $\pi$ range, see solid curve in Fig.~\ref{fig:kappaz=0}.
Given an initial angular momentum only in $x$ direction, the molecular axis
then continuously rotates around the $x$ axis so that while $L_x$ is modulated
due to the potential barriers, it does not change its sign, see solid curve in
Fig.~\ref{fig:Ly-kappaz=0}.

On the other hand, for the same case of $\kappa_z \!=\! 0$, for energies
$\epsilon \!<\! 0$ the angle $\vartheta$ is subject to a reflection at the
light-induced potential barrier. Thus the range of $\vartheta$ values is
restricted, as seen for the dashed line in Fig.~\ref{fig:kappaz=0}. In
addition, since $\dot\vartheta$ changes its sign at the reflecting barrier,
the angular momentum $L_x$ correspondingly crosses zero and thus rotations
with positive and negative orientations subsequently interchange, cf. dashed
line in Fig.~\ref{fig:Ly-kappaz=0}. Note that for $\kappa_z \!>\!  0$ the
motion of $\vartheta$ is additionally bound by reflections at the centrifugal
barriers near $\vartheta \!=\! 0$ and $\pi$.

\section{Quantum dynamics of the molecular rotation} \label{sec:3}

\subsection{Implementation via two-photon Raman transitions}
\begin{figure}
  \begin{center}
   \epsfig{file=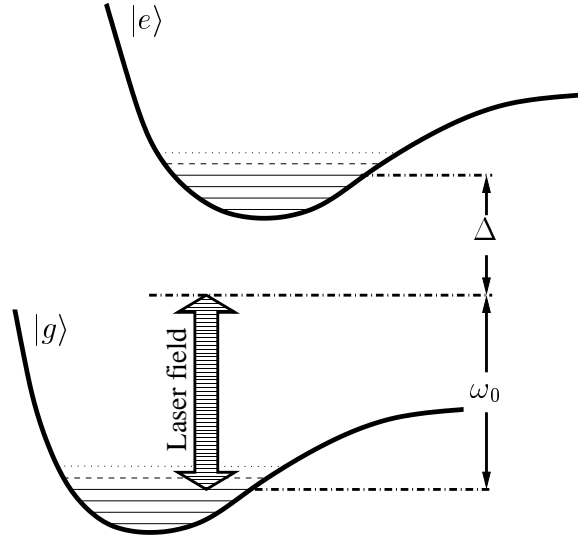,scale=1} 
  \end{center}
  \caption{Excitation scheme for implementing rotational Raman transitions.}
  \label{fig:level-scheme}
\end{figure}

For implementing the dynamics described so far, not only static electric
fields can be used. Also linear polarised cw optical fields, that are far
detuned from electronic resonances, have been shown to produce a similar
interaction~\cite{herschbach-pol}. In this way one can take advantage of large
polarisabilities of non-polar molecules to attain deep potential wells for the
nutation angle. The specific excitation scheme for a dimer, depicting only the
two lowest electronic potential surfaces that provide a
$\leftexp{1}{\Sigma}\!\leftrightarrow\!  \leftexp{1}\Sigma$ dipole transition,
is shown in Fig.~\ref{fig:level-scheme}.

In order to avoid resonant vibronic transitions and to be left with the sought
two-photon processes, a sufficiently large detuning $\Delta$ from the bare
electronic resonance frequency is required. Given an optical field of spectral
width much smaller than vibrational frequencies $\omega_\nu$, stimulated Raman
transitions occur that only affect the rotational degree of freedom. However,
the relative rate of spontaneous vs stimulated Raman transitions,
$\Gamma/\Delta$, with $\Gamma$ being the bare electronic linewidth of the
dipole transition, shows that even for large detuning, $\Delta \!\gg\!
\Gamma$, a certain number of spontaneous processes occur. These processes not
only lead to an incoherent effect on the molecular rotation, such as
relaxation and quantum decoherence, but also excite the internuclear vibration
via Franck--Condon transitions.

Nevertheless, as long as the laser remains detuned from resonant vibronic
transitions, the electronic and vibrational degrees of freedom can be
eliminated to obtain equations of motion for the molecular rotation
alone.
%WAL:
This is achieved by consistently eliminating the electronic coherences
and electronic excited-state populations, keeping only terms up to
second order in the molecule-light interaction. In Franck--Condon
approximation the vibrational degree of freedom can then be traced
over~\cite{mol-rot}.
%This 
The requirement 
of avoiding resonant vibronic transitions
leads to a maximum interaction time on which the molecule stays
vibrationally sufficiently cold, which can be estimated as
\begin{equation}
  t \!\ll\! \Delta^2 / (\eta^2 \Gamma \omega_\nu
\Omega_{\rm R}) .
\end{equation}
Here $\Omega_{\rm R} \!=\!  |d_{\rm ge}E_0|^2 / (4 \hbar^2 \Delta)$ is the
Raman Rabi frequency with $d_{\rm ge}$ being the electric-dipole matrix
element between electronic ground and excited states.  The linear polarised
electric field is specified by $E(t) \!=\! \hat{n}_z \, E_0 \cos(\omega_0 t)$
with $\omega_0$ being an optical frequency far-detuned from vibronic resonance
and $\hat{n}_z$ being the unit vector in $z$ direction. Using these parameters
the light-induced potential depth results as $U_\alpha \!=\! \hbar\Omega_{\rm
  R}$.

The parameter $\eta$ is the ratio of the difference of internuclear
equilibrium distances in ground and excited electronic states to the spatial
extent of the vibrational ground-state wave packet.  Since $\eta \!<\! 10$ for
alkali dimers, in the weak-field regime [$\Omega_R \!<\! \hbar /(2\Theta)$]
this time corresponds to a very large number of free rotational periods.

Under these conditions the equations of motion for the orientation of the
molecular axis in the far detuned optical field can be cast into a master
equation of Lindblad form~\cite{mol-rot}:
\begin{eqnarray}
  \label{eq:master}
  \dot{\hat{\sigma}} = - 
    \frac{i}{\hbar} [ \hat{T} \!+\! \hat{V} , \hat{\sigma} ] + \sum_{i=x,y,z}
    \left(
    \hat{S}_i \, \hat{\sigma} \,  
  \hat{S}^\dagger_i - {\textstyle\frac{1}{2}} \, \hat{S}^\dagger_i 
  \hat{S}_i \, \hat{\sigma} - {\textstyle\frac{1}{2}} \, \hat{\sigma}
    \, \hat{S}^\dagger_i \hat{S}_i  \right) .
\end{eqnarray}
Here $\hat{\sigma}$ is the reduced rotational density operator of the molecule
and the free kinetic energy is given by~\cite{LJ-remark}
\begin{equation}
  \label{eq:ham-mol}
  \hat{T} = \hat{\bf J}^2 / (2\Theta) .
\end{equation}
The stimulated rotational Raman transitions then provide the sought potential
in the form
\begin{equation}
  \label{eq:raman-ham}
  \hat{V} = - \hbar\Omega_{\rm R} \, \hat{n}_z^2 ,   
\end{equation}
where in terms of spherical angles the unit-vector operator $\hat{n}_z$ reads
$\langle \vartheta, \varphi| \hat{n}_z \!=\! \cos\vartheta \, \langle
\vartheta, \varphi|$, so that
\begin{equation}
  \label{eq:raman-ham2}
  \langle \vartheta, \varphi | \, \hat{V} = 
  - U_\alpha \cos^2(\vartheta) \, 
  \langle \vartheta, \varphi| 
\end{equation}
recovers the potential as discussed for the classical treatment. 
 
The spontaneous rotational Raman transitions are given in terms of the
operators $\hat{S}_i$, with $i \!=\! x,y,z$ denoting the polarisation of the
spontaneously emitted photon. They read
\begin{equation}
  \label{eq:def-S}
  \hat{S}_i = \sqrt{\frac{\Gamma \Omega_{\rm R}}{\Delta}} \; 
  \hat{n}_i \, \hat{n}_z , 
\end{equation}
with $\hat{n}_i$ ($i \!=\! x,y,z$) being unit vectors pointing into the
corresponding directions.  In the representation of spherical angles these
jump operators can be cast in vector form to:
\begin{equation}
  \label{eq:def-S2}
  \langle \vartheta, \varphi| \, \hat{\vec{S}} 
  = \sqrt{\frac{\Gamma \Omega_{\rm R}}{\Delta}} \left[
  \begin{array}{c} 
    \cos\varphi \sin\vartheta \cos\vartheta \\
    \sin\varphi \sin\vartheta \cos\vartheta \\
    \cos^2\vartheta
  \end{array} \right] \langle \vartheta, \varphi| .
\end{equation}

\subsection{Stimulated Raman processes}
\begin{figure}
  \begin{center}
    \epsfig{file=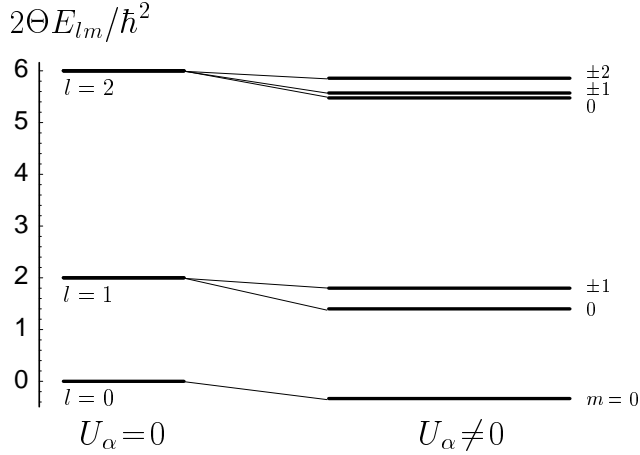,scale=0.75}
  \end{center}
  \caption{Change of energy eigenvalues due to the light interaction for
    the value $2\Theta U_\alpha / \hbar^2 \!=\! 1$.}
  \label{fig:E-levels}
\end{figure}

For vanishing bare electronic linewidth, $\Gamma \!=\! 0$, a unitary evolution
is recovered from the master equation~(\ref{eq:master}) that can be
reformulated in terms of the Schr\"odinger equation
\begin{equation}
  \label{eq:schroedinger}
  i \hbar \, \partial_t \, |\Psi\rangle = (\hat{T}  \!+\! \hat{V} ) \, 
  |\Psi\rangle , 
\end{equation}
with the Hamiltonian given by Eqs~(\ref{eq:ham-mol}) and
(\ref{eq:raman-ham}). In the $|\vartheta, \varphi\rangle$ representation the
evolution of the wavefunction $\Psi(\vartheta,\varphi) \!=\! \langle
\vartheta, \varphi| \Psi\rangle$ reads
\begin{equation}
  \label{eq:schroedinger2}
  \fl i \hbar \, \dot{\Psi}(\vartheta,\varphi) = - \left\{
    \frac{\hbar^2}{2\Theta} 
    \left[ \frac{1}{\sin\vartheta} \partial_\vartheta \left( \sin\vartheta
        \, \partial_\vartheta \right) + \frac{1}{\sin^2\vartheta} \,
      \partial_\varphi^2 \right] + U_\alpha 
    \cos^2\vartheta \right\} \, \Psi(\vartheta,
  \varphi) .
\end{equation}
Using the ansatz $\Psi(\vartheta, \varphi) \!=\! S_{lm}(z) \exp(im\varphi)$,
with $\hbar m \!=\! J_z$ being a constant of motion, and $z \!=\!
\cos\vartheta$, the energy eigenstates are obtained from the angular oblate
spheroidal wave equation~\cite{as}
\begin{equation}
  \label{eq:schroedinger2}
  \left\{ \frac{\hbar^2}{2\Theta}
  \left[ \partial_z \left[ (1 \!-\! z^2) \partial_z \right]- \frac{m^2}{1
  \!-\! z^2} \right] + U_\alpha z^2 + E_{lm} \right\} S_{lm}(z) = 0 . 
\end{equation}
The corresponding eigenenergies $E_{lm}$ can be expanded in powers of the
interaction potential $U_\alpha$ as
\begin{equation}
  \label{eq:energies}
  E_{lm} = \frac{\hbar^2}{2\Theta} \, l(l \!+\! 1) - \frac{U_\alpha}{2} 
  \left[ 1 + \frac{1 \!-\! 4m^2}{(2l \!-\! 1) (2l \!+\! 3)}
  \right] + \ldots \, .
\end{equation}
They manifest a partial removal of the degeneracy with respect to $m$, as
shown in Fig.~\ref{fig:E-levels}. The properties of these eigenstates and
their angular confinement have been discussed in detail in
Ref.~\cite{herschbach-E}.

\begin{figure}
  \centering
  \epsfig{file=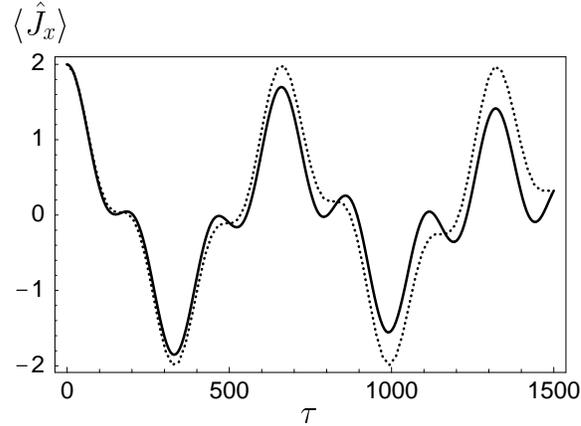,scale=0.75}
  \caption{Dynamics of $\langle \hat{J}_x(t) \rangle$ over scaled,
    dimensionless time $\tau \!=\! t \hbar / (2\Theta)$. The laser interaction
    strength is $\hbar\Omega_{R} / (2\Theta) \!=\! 0.1$ and the detuning
    $\Gamma/\Delta \!=\! 0.01$ (solid line). The dotted line corresponds to
    $\Gamma \!= \!0$.}
  \label{fig:Lx-dynamics}
\end{figure}

The quantum dynamics of a system prepared quite analogously as discussed for
the classical treatment, i.e. with an initial average angular momentum
pointing only in $x$ direction, is shown in Fig.~\ref{fig:Lx-dynamics}. Here
an initial coherent angular-momentum state~\cite{Arecchi} of the form
\begin{equation}
  \label{eq:coh-state}
  \fl | j, \alpha, \beta \rangle = \sum_{m=-j}^j  \sqrt{ 2j \choose j \!+\! m
   } \big[\sin (\alpha/2) \big]^{j+m}   \big[\cos
  (\alpha/2) \big]^{j-m} \, e^{-i(j + 
     m)\beta} |jm\rangle , 
\end{equation}
for $j \!=\! 2$, $\alpha \!=\! \pi/2$, and $\beta \!=\! 0$ has been used.
Classically that would correspond to the case $\kappa_z \!=\! 0$ where
reflections in the motion of $\vartheta$ can only occur due to the
light-induced potential. But from Eq.~(\ref{eq:coh-state}) it is clear that we
deal with an initial state containing states $|j,m\rangle$ with all possible
values for $m$:
\begin{eqnarray}
  \label{eq:coh-state2}
  |\Psi_1\rangle & = & 
  |j\!=\! 2 ,\alpha \!=\! \pi/2, \beta \!=\!0\rangle \nonumber \\
  & = & 
  {\textstyle\frac{1}{4}} \, |2,-2\rangle +
  {\textstyle\frac{1}{2}} \, |2,-1\rangle + {\textstyle\sqrt{\frac{3}{8}}} \, |2,0\rangle + 
  {\textstyle\frac{1}{2}} \, |2,1\rangle + {\textstyle\frac{1}{4}} \,
  |2,2\rangle  .  
\end{eqnarray}
Thus part of the wave packet in $\vartheta$ will also be reflected by
centrifugal barriers.

In Fig.~\ref{fig:Lx-dynamics} an oscillation of $\langle \hat{J}_x \rangle$
between positive and negative values is observed that is characteristic for
reflections of the wavefunction at potential barriers.  However, the transient
behaviour at the crossing of zero indicates that part of the wave packet in
$\vartheta$ is transmitted through the barrier, and that reflected and
transmitted parts interfere for the short time where they spatially overlap.

\subsection{Spontaneous processes}
When spontaneous processes are included in the dynamics, i.e for $\Gamma \!>\!
0$, the full master equation~(\ref{eq:master}) must be solved. Due to the
scaling of complexity of this numerical problem with the number of required
rotational levels, it is advantageous to employ a quantum trajectory method.
Then individual realisations of state vectors can be calculated and finally an
ensemble average over the set of realisation reproduces the sought density
matrix.

Individual trajectories are comprised of non-unitary evolutions with the
effective Hamiltonian
\begin{equation}
  \label{eq:Heff}
  \hat{H}_{\rm eff} = \hat{T} + \hat{V} \left( 1 + \frac{i \Gamma}{2\Delta}
  \right) , 
\end{equation}
describing the conditioned evolution when no spontaneous processes occur,
intermitted by spontaneous Raman processes described by the application of the
jump operators $\hat{S}_i$. As can be seen from their
definition~(\ref{eq:def-S2}), these operators may excite the motion in the
angle $\varphi$, by producing via $\hat{S}_x$ and $\hat{S}_y$ a weighting of
the corresponding probability amplitude $\Psi(\vartheta,\varphi)$ in
$\varphi$. Thus motion in $\varphi$ and consequently the angular momentum
component $\hat{J}_z$ are getting excited by spontaneous Raman processes.

\begin{figure}
  \centering
  \epsfig{file=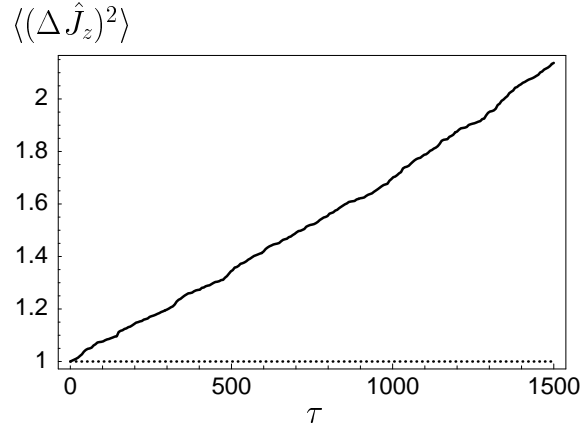,scale=0.75}
  \caption{Variance of $\hat{J}_z$ over scaled, dimensionless time,
    $\tau \!=\! t \hbar/ (2\Theta)$. The parameters are the same as in
    Fig.~\ref{fig:Lx-dynamics}. }
  \label{fig:Lz2}
\end{figure}

This can be observed in Fig.~\ref{fig:Lz2}, where the variance of $\hat{J}_z$
is plotted over time. For $\Gamma \!>\! 0$ a monotonous increase of the
variance is observed, indicating that the kinetic energy is subject to
heating.  Of course the spontaneous Raman scattering will also lead to a
suppression of coherent processes and the oscillations of the average angular
momentum will be damped, cf. solid curve in Fig.~\ref{fig:Lx-dynamics}.

\section{Decohering dynamics in phase space} 
\label{sec:4} 

\subsection{Spherical Wigner functions}
In the same way as the angular-momentum coherent states are a generalisation
of the familiar harmonic-oscillator coherent states, also the concept of
phase-space quasi-probability distributions can be defined on the
angular-momentum phase space~\cite{Agarwal}. Here we employ a Wigner function,
which in the $(2j \!+\! 1)$-dimensional Hilbert space of angular momentum $j$,
is defined in terms of the density operator $\hat{\sigma}$ and the spherical
harmonics, $Y_{sm}(\theta, \varphi)$, as defined in
Refs~\cite{defn-w:Schleich, defn-w:Benedict}
\begin{eqnarray}
  \label{eq:W-fn}
  W_j(\vartheta, \varphi)  & = & \sqrt{\frac{2j\!+\!1}{4\pi}} \sum_{s=0}^{2j}
 \sum_{m=-s}^s Y_{sm}(\vartheta, \varphi) \, {\rm Tr}(\hat{T}_{j,sm}^\dagger 
 \hat{\sigma} ) . 
\end{eqnarray}
The multipole operators are defined by
\begin{eqnarray}
  \label{eq:multipole}
  \hat{T}_{j,sm} & = & \sqrt{2s \!+\! 1}  \sum_{m'm''} (-1)^{j-m'} 
  \left( \begin{array}{ccc}
      j & s & j \\
      -m' & m & m''
      \end{array} \right) |j m' \rangle \langle j m''| ,
\end{eqnarray}
with ${j_1 \, j_2 \, j_3 \choose m_1 \, m_2 \, m_3}$ being the Wigner 3j
symbol. The factor in front of Eq.~(\ref{eq:W-fn}) ensures normalisation to
unity.

As the laser interaction will result in a change of the rotational quantum
number $j$ of the molecule we consider a sum of Wigner functions over all
possible values of $j$:
\begin{equation}
  \label{eq:wigner}
  W(\vartheta, \varphi) = \sum_{j=0}^\infty W_j(\vartheta, \varphi) .
\end{equation}
%WAL:
Albeit the sum over $j$, this phase-space distribution is not a
complete description of the rotational quantum state. However, it
proves useful for illustrating the laser-induced dynamics. For
alternative approaches defining phase-space distributions independent
of the particular value of $j$, see Refs~\cite{foeldi,brif,klimov}.
Moreover,
it should be noted that even though the angular-momentum coherent
states are analogous to the harmonic-oscillator ones in many respects,
the Wigner function of angular-momentum coherent states is weakly
negative. This negativity decreases with increasing value for $j$ and
reaches zero only in the (classical) limit $j\!\to\!\infty$.

\subsection{Time evolution of the Wigner function}

\begin{figure}
  \begin{center}
    \epsfig{file=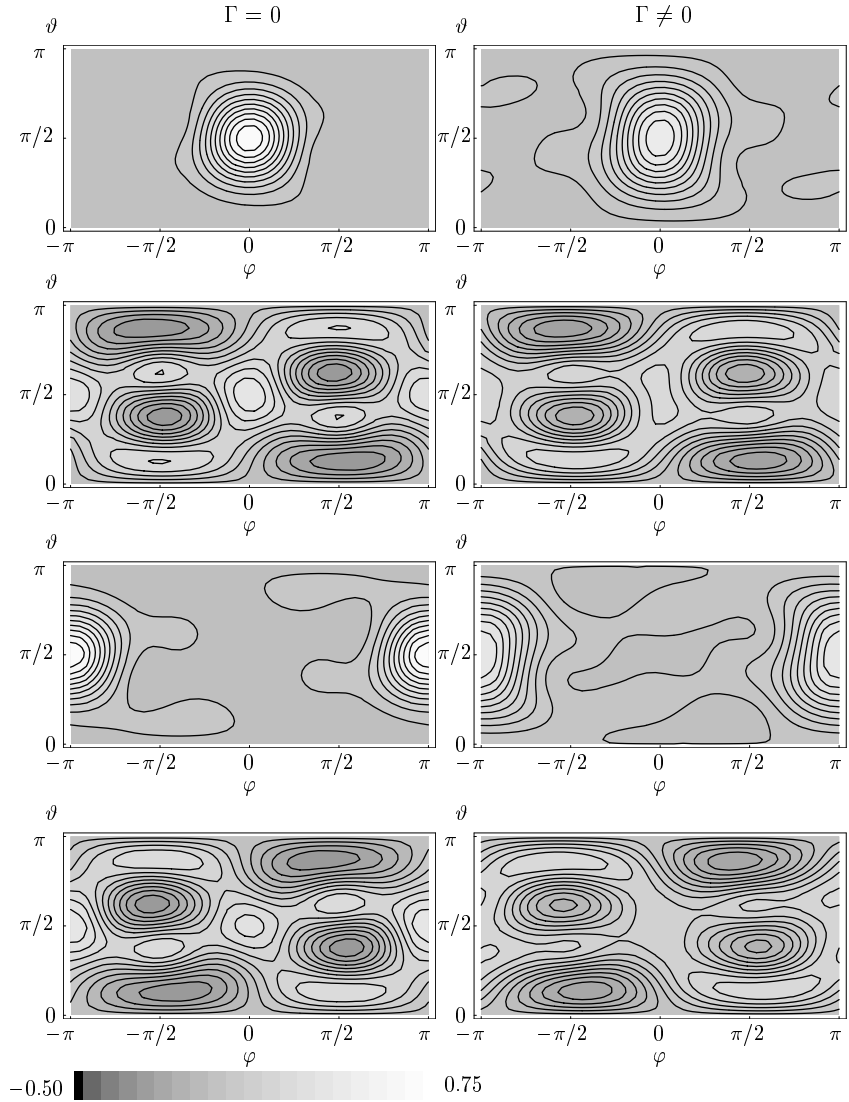,scale=0.75}
  \end{center}
  \caption{Time evolution of the coherent state $|\Psi_1\rangle$. From
    top to bottom the figures show the Wigner function at scaled times $\tau
    \!=\! 660$, $\tau \!=\! 830$, $\tau \!=\! 990$, and $\tau \!=\! 1165$.
    Parameters are the same as in Fig.~\ref{fig:Lx-dynamics} and $\tau \!=\!
    t\hbar/(2 \Theta)$.}
  \label{fig:W-fn_coh-state}
\end{figure}

Figure~\ref{fig:W-fn_coh-state} shows the time evolution of the system
starting initially from the state $|\Psi_1\rangle$,
cf.~Eq.~(\ref{eq:coh-state2}). The phase-space distribution is shown for a
scaled time $\tau \!=\! t\hbar/(2 \Theta) \!=\! 660$ after the first full
period of oscillation of $\langle \hat{J}_x \rangle$ (upper left plot), and
then at later times in steps of quarter periods.  One can see that the system
almost returns to its initial coherent state after one period, when the
spontaneous emission is neglected (left column).  The peak in phase space has
then returned to its initial position at $\vartheta \!=\! \pi/2$ and $\varphi
\!=\!  0$. In the course of time the peak splits and after a half period, at
$\tau=990$, two peaks appear at $\vartheta \!=\! \pi/2$, $\varphi \!=\!
0,\pi$. That corresponds to an average angular momentum pointing now in
negative $x$ direction. Note that between half and full cycles, at $\tau \!=\!
830$ and $\tau \!=\! 1165$, the phase-space distribution develops several
smaller peaks and also negative regions of the Wigner function appear.
Including the spontaneous Raman processes, the structures in phase-space
generally are smoothed and negativities are suppressed. This can be seen from
the right column of plots in Fig.~\ref{fig:W-fn_coh-state}.

\begin{figure}
  \begin{center}
    \epsfig{file=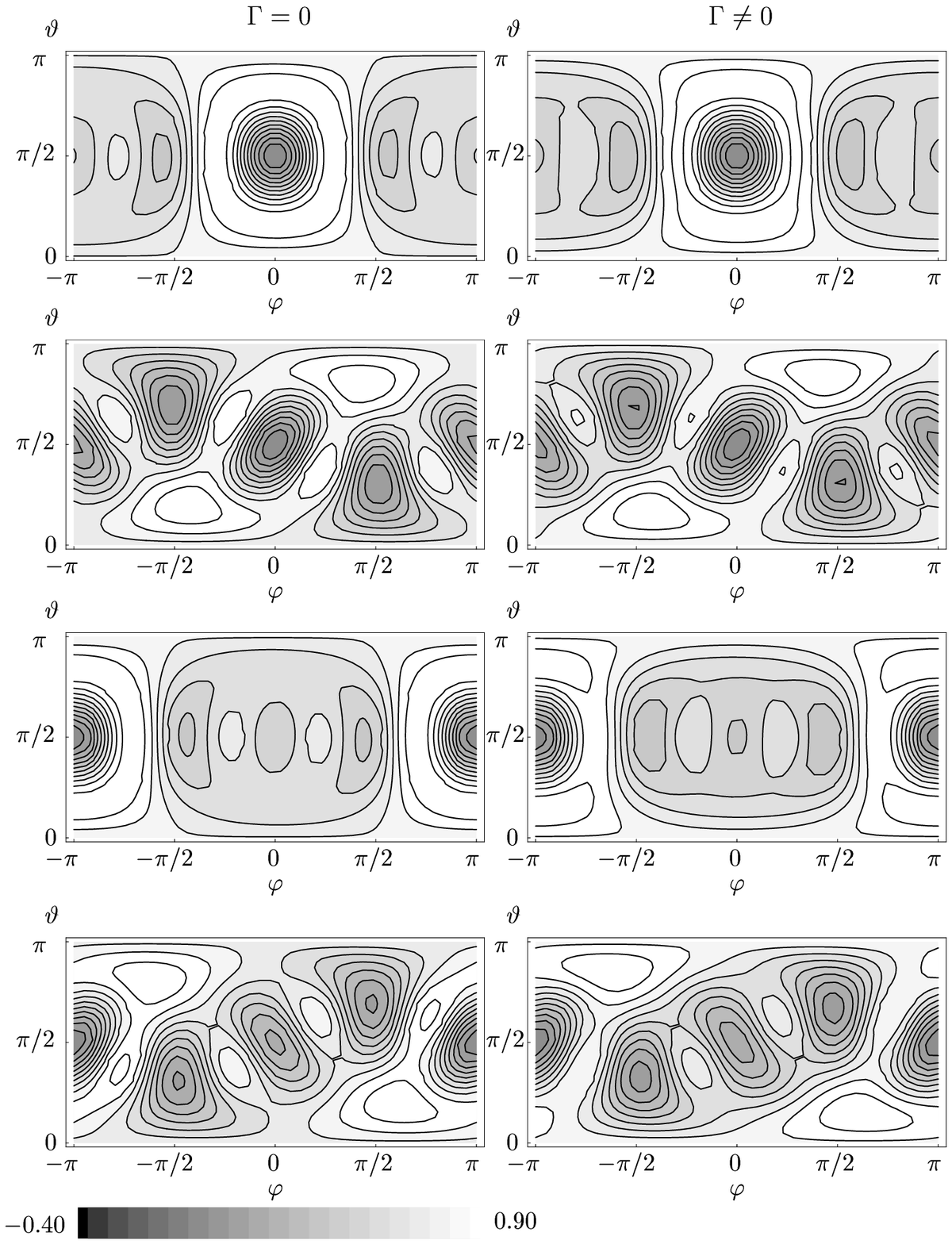,scale=0.75}
  \end{center}
  \caption{Time evolution of the initial superposition state $|\Psi_2\rangle$. 
    Parameters are the same as in Fig.~\ref{fig:W-fn_coh-state} and $\tau
    \!=\!  t\hbar / (2\Theta)$.}
  \label{fig:W-fn_sup-state}
\end{figure}

As another example, in Fig.~\ref{fig:W-fn_sup-state} a superposition of two
coherent angular-momentum states,
\begin{equation}
  \label{eq:sup-state}
  |\Psi_2 \rangle = \sqrt{\frac{2}{5}} \Big( \, |
  j \!=\! 2, \alpha \!=\! \pi/2, \beta \!=\! \pi/4 \rangle + | j \!=\! 2,
  \alpha \!=\! \pi/2, \beta \!=\! -\pi/4 \rangle \Big) , 
\end{equation}
has been taken as the initial quantum state. This state has a strong negative
region of the Wigner function centred at $\vartheta \!=\!  \pi/2$, $\varphi
\!=\! 0$.  The plots of the Wigner function for this initial state show the
same oscillatory behaviour as Fig.~\ref{fig:W-fn_coh-state}, with the negative
peak being shifted to $\varphi \!=\! 0,\pi$ at half periods.

\begin{figure}
  \begin{center}
    \epsfig{file=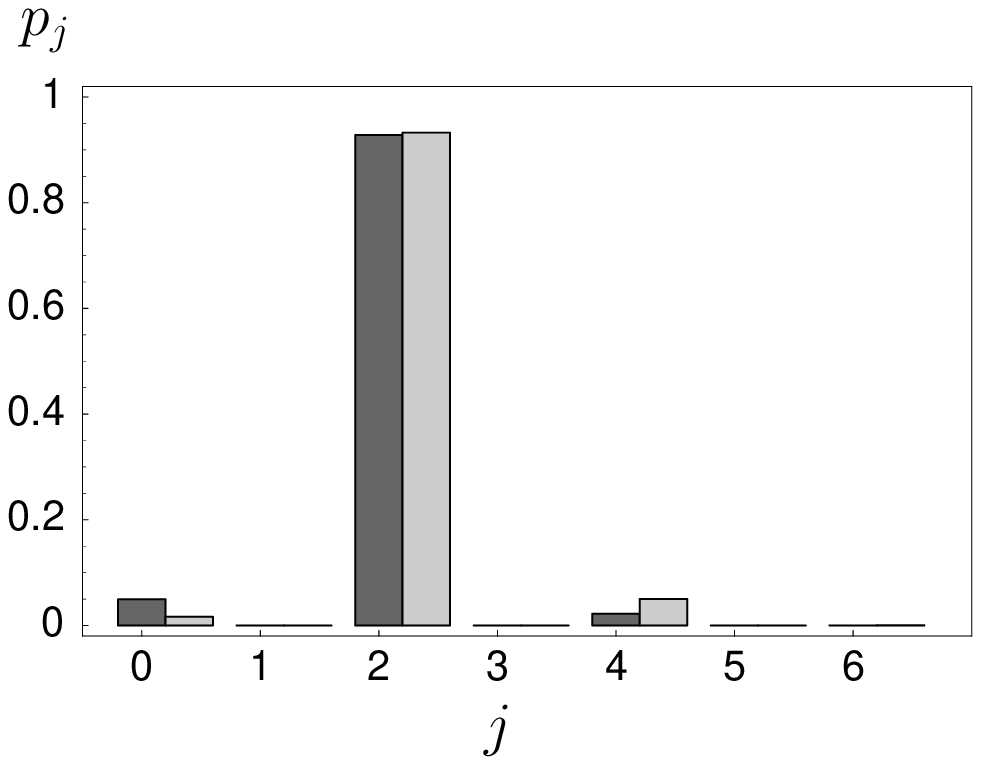,scale=0.75}
  \end{center}
  \caption{Populations $p_j$ at the interaction time $\tau \!=\!
    1165$ for the coherent angular-momentum state (light bars) and the
    superposition of coherent angular-momentum states (dark bars).
    Parameters are the same as for Figs~\ref{fig:W-fn_coh-state},
    \ref{fig:W-fn_sup-state}.}
  \label{fig:jpops}
\end{figure}

However, comparing the time evolution of these two cases one can see that the
difference between the unitary and non-unitary evolution is much more
pronounced in the case of the coherent state. Especially at the times between
completed half and full cycles (at $\tau \!=\! 830$ and $\tau \!=\! 1165$),
when also the Wigner function of the time-evolved coherent state shows
relative strong negative regions, one observes that these negativities have
become weaker in the presence of spontaneous emission. 
%WAL:
The populations in the $j$-level manifolds for the final interaction
time of Figs~\ref{fig:W-fn_coh-state} and \ref{fig:W-fn_sup-state} are
shown in Fig.~\ref{fig:jpops}. It can be seen that for both the
coherent state and the superposition of coherent states the
populations in $j \!=\!  0,4$ have increased by a few percent at the
expense of the population in the initial manifold $j \!=\! 2$.

That the coherent state in this example actually is much more sensitive to
decoherence may seem counter-intuitive at first sight, but one should keep in
mind that the angular-momentum coherent state is not classical in the usual
sense of a harmonic-oscillator coherent state. That is, as has been seen above
its Wigner function in general is not positive everywhere. Furthermore,
angular-momentum coherent states apparently are not in general the most robust
states against decoherence, as are harmonic oscillator coherent states for
linearly coupled reservoirs.

\begin{figure}
  \begin{center}
    \epsfig{file=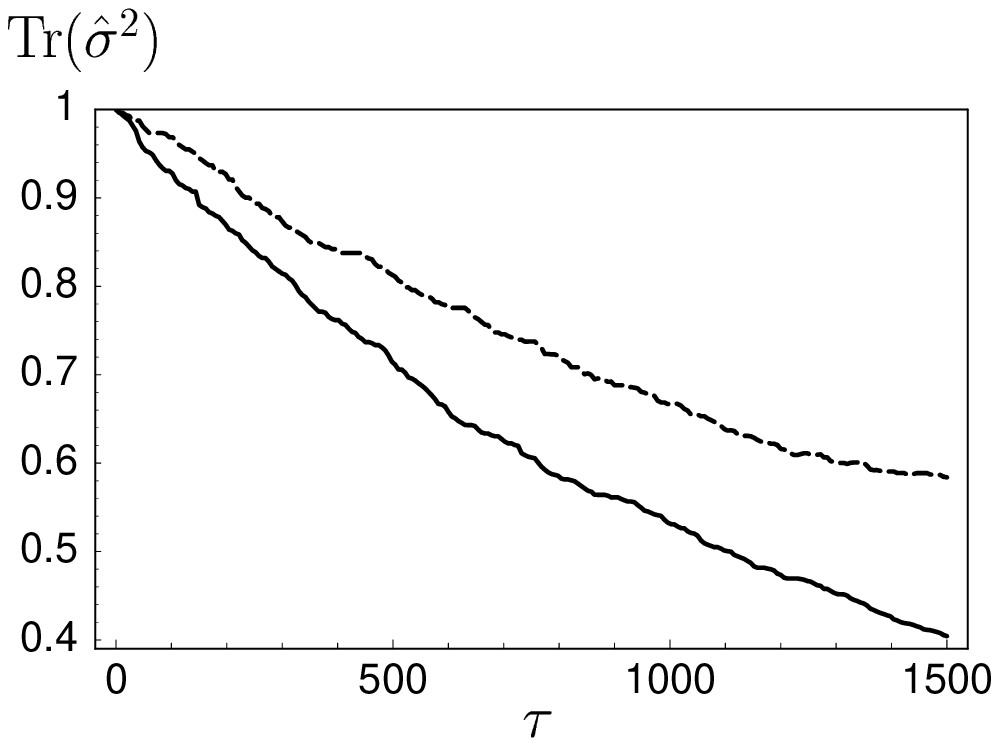,scale=0.75}
  \end{center}
  \caption{Purity for the initial coherent state (solid curve) and initial
    superposition state (dashed curve). Parameters are the same as in
    Fig.~\ref{fig:Lx-dynamics} and $\tau \!=\! t\hbar / (2\Theta)$.}.
  \label{fig:purity}
\end{figure}

The substantial difference in how these two cases are affected by decoherence
due to spontaneous processes can also be observed in Fig.~\ref{fig:purity},
where the time evolution of the purity of the density operator is shown.  It
can be seen that the superposition state~(\ref{eq:sup-state}) develops much
slower into a mixed state compared to the coherent state. 

%WAL:
The type of decoherence mechanism is in general determined by the form
of the operators $\hat{S}_i$, see Eqs~(\ref{eq:def-S}) and
(\ref{eq:def-S2}). In our case these operators differ from raising and
lowering operators $\hat{J}_\pm \!=\!  \hat{J}_x \!\pm\! i \hat{J}_y$,
and thus our decoherence mechanism itself is different from that
discussed in Refs~\cite{braun,foeldi2}, though for special choices of
initial conditions particular features may be similar.

\section{Summary and conclusions}
We have considered a diatomic homonuclear molecule in a far-detuned,
linear-polarised light field. First the classical motion of the molecule was
studied. For the case of a molecule with a permanent dipole, the potential was
seen to be that of a spherical pendulum. For non-polar molecules, however, the
potential is of a different form, but the solution for the nutation angle
still could be found in terms of elliptic integrals.  The motion was then
treated quantum mechanically, where stimulated and spontaneous Raman processes
were taken into account by a master equation. Solving this equation both for
an angular-momentum coherent state and a superposition state, the effects of
the decoherence induced by the spontaneous processes where compared.  It was
seen that, in this particular example, the superposition state was more robust
to decoherence than the coherent state.

\section*{Acknowledgements}
This research was supported by Deutsche Forschungsgemeinschaft.


\begin{thebibliography}{99}
\bibitem{LJ-remark} For notational correctness we use in the quantum case the
  angular momentum operator $\hat{\vec{J}}$ instead of $\hat{\vec{L}}$.
\end{thebibliography}
\end{document}